# REVIEW

# So You Want to Image Myelin Using MRI: Magnetic Susceptibility Source Separation for Myelin Imaging

Jongho Lee[1*] , Sooyeon Ji[1] , and Se-Hong Oh[2*]

In MRI, researchers have long endeavored to effectively visualize myelin distribution in the brain, a pursuit with significant implications for both scientific research and clinical applications. Over time, various methods such as myelin water imaging, magnetization transfer imaging, and relaxometric imaging have been developed, each carrying distinct advantages and limitations. Recently, an innovative technique named as magnetic susceptibility source separation has emerged, introducing a novel surrogate biomarker for myelin in the form of a diamagnetic susceptibility map. This paper comprehensively reviews this cutting-edge method, providing the fundamental concepts of magnetic susceptibility, susceptibility imaging, and the validation of the diamagnetic susceptibility map as a myelin biomarker that indirectly measures myelin content. Additionally, the paper explores essential aspects of data acquisition and processing, offering practical insights for readers. A comparison with established myelin imaging methods is also presented, and both current and prospective clinical and scientific applications are discussed to provide a holistic understanding of the technique. This work aims to serve as a foundational resource for newcomers entering this dynamic and rapidly expanding field.

## Introduction

The central nervous system, an intricate network of neurons and glia, relies on a pivotal player – myelin – to orchestrate rapid and precise neural communication. Myelin, composed of a lipid bilayer membrane, envelops axons, forming a protective sheath. Its high cholesterol content, which reflects and scatters light, contributes to the distinct white appearance of the brain's white matter. The functional significance of myelin lies in its ability to insulate nerve fibers, preventing signal loss and facilitating saltatory conduction. In MRI, the distinct properties of myelin make it a key determinant in multiple MRI contrasts observed between white and gray matter (e.g., $T_1$-weighted image, $T_2$-weighted image, and susceptibility), offering invaluable insights into the brain structure, function, and pathology.

Because of its value in clinic and neuroscience, the pursuit of myelin-specific imaging methods has been a long-standing objective in MRI. Various techniques, including myelin water imaging (MWI),[1–4] magnetization transfer imaging and its variations,[5–8] diffusion imaging,[9] myelin volume fraction from synthetic MRI,[10,11] and other relaxometry imaging,[11–18] have been developed as biomarkers that provide sensitivity and specificity to myelin. More recently, magnetic susceptibility imaging[19,20] has been proposed, leveraging the diamagnetic susceptibility characteristics of myelin.[21,22] While each method aims to serve as a biomarker for myelin, they come with inherent advantages and limitations.[23–28] Consequently, the field is characterized by ongoing efforts to develop novel contrast mechanisms and refine existing methods, reflecting the continuous quest for myelin-specific imaging.

These myelin imaging methods have found a number of applications in neuroscience and clinical research. For example, they have been utilized to create the myeloarchitecture of the neocortex,[13,16,29–35] which has an important value in cortical parcellation. The studies of myelin change during development[36–46] and normal aging[47–49] have also been conducted using myelin imaging, revealing age-dependent myelin concentration changes. More recently, the methods have been applied to the studies of brain plasticity, suggesting training-induced myelin changes.[50,51] In clinical research, the techniques have been applied to various neurological disorders including multiple sclerosis (MS)[52–57] and leukodystrophy.[58–61] In particular, MS, where loss of myelin is the hallmark of the disease, has been the target of the myelin imaging methods, exploring diagnosis,[62–64] lesion characterization,[52,53,65–68] disease progression monitoring,[69–71] and treatment assessment.[72]

[1]Department of Electrical and Computer Engineering, Seoul National University, Seoul, Korea
[2]Biomedical Engineering, Hankuk University of Foreign Studies, Yongin, Korea
Jongho Lee and Sooyeon Ji contributed equally to the manuscript.
*Corresponding authors: Jongho Lee, PhD, Department of Electrical and Computer Engineering, Seoul National University, Seoul, Republic of Korea. Phone: +82-2-880-7310, E-mail: jonghoyi@snu.ac.kr
Se-Hong Oh, PhD, Department of Biomedical Engineering, Hankuk University of Foreign Studies, Yongin, Republic of Korea. Phone: +82-31-330-4592, E-mail: jakeoh79@gmail.com

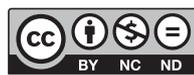

In this review, we will delve into the technique of advanced magnetic susceptibility imaging for myelin imaging.[73] This relatively new field, coined as magnetic susceptibility source separation,[35,64,73–84] presents a novel avenue for generating an indirect measure of myelin information in both gray and white matter of the brain, overcoming some of the limitations associated with traditional myelin-specific imaging approaches. By employing an advanced biophysical model and leveraging the diamagnetic susceptibility characteristics of myelin, the susceptibility source separation aims to provide high-resolution quantitative myelin information. In this review, we navigate the landscape of this emerging technique, exploring its fundamental physics and elucidating both advantages and challenges inherent in this approach. Our scope extends to considerations of data acquisition and processing, providing commentary on the current and potential applications of this novel method in the quest to map myelin distribution in the brain. The structure of this review is formed in nine questions and answers with some overlap among the answers. Finally, we hope you have an opportunity to gain an overview of this emerging and rapidly growing technique of magnetic susceptibility source separation.

## Questions and Answers

### Question 1: Let me begin with basic questions. What is susceptibility, and how does it affect MRI?

**Answer 1:** Magnetic susceptibility is the degree to which a material is magnetized in response to an applied magnetic field. This can be expressed as $B = \mu H = \mu_0(1+\chi)H$ for a linear and isotropic material, where $B$ is the magnetic field experienced by the material, $\mu$ is the magnetic permeability of the material, $\mu_0$ is the magnetic permeability of vacuum, $\chi$ is the magnetic susceptibility of the material, and $H$ is the applied field. For example, if a sphere of deoxygenated hemoglobin proteins is placed in water in a magnetic field (H-field), the sphere will experience a higher field than water (B-field) because the deoxygenated hemoglobin has higher susceptibility ($\chi = +0.15$ ppm)[85] than that of water ($\chi = -9.05$ ppm)[85] (Fig. 1a). Note that the field change is not confined to the sphere. Depending on the sign of $\chi$, we categorize a material as diamagnetic if $\chi$ is negative (e.g., cholesterol: $\chi = -9.23$ ppm),[86] paramagnetic if $\chi$ is small positive (e.g., ferritin: $\chi = 520$ ppm),[85] or ferromagnetic if $\chi$ is large positive. In MRI, we do not scan ferromagnetic materials due to safety and artifacts,[85] limiting the materials of interest to para- and diamagnetic materials. Although this categorization relative to the susceptibility of vacuum ($\chi = 0$ ppm) is universal, many MRI literature uses water susceptibility as the reference.[87,88] We will also use water as the reference for the susceptibility maps of this review.

In MRI, susceptibility has long been an enemy and a friend. It not only gives rise to artifacts[89,90] (Fig. 1b[91] and 1c[90]) but also serves as a cornerstone for advanced imaging techniques including functional MRI (fMRI),[92] susceptibility weighted imaging (SWI)[93,94] (Fig. 1c[90]), and quantitative susceptibility mapping (QSM)[19,20,95,96] (Fig. 1d[97]). Susceptibility-induced artifacts, largely due to the susceptibility difference between air and water, are manifested by geometric distortions in echo-planar imaging (blue arrow in Fig. 1b) and/or signal loss in SWI (red arrow in Fig. 1c), particularly near the nasal cavity and ear canal, creating challenges in MRI. On the other hand, the temporal susceptibility change from the concentration variation in deoxygenated hemoglobin due to neurovascular coupling is the primary contrast mechanism for fMRI, whereas the spatial susceptibility variation of tissue iron (e.g., ferritin and hemosiderin) and deoxygenated hemoglobin is the contrast source for SWI. In QSM, both paramagnetic sources (e.g., ferritin, hemosiderin, and deoxyhemoglobin) and diamagnetic sources (e.g., myelin and calcium) are responsible for the image contrast, delineating details of brain structures that may not be visible in conventional MRI methods.

### Question 2: Susceptibility imaging is new to me. Can you give a brief introduction?

**Answer 2:** Sure. As I mentioned in Question and Answer 1, magnetic susceptibility serves as an important contrast source for MRI, and there exist a few well-known susceptibility imaging methods such as SWI and QSM that provide important information on tissue microstructure and composition.

### SWI

One of the earliest methods in this field is SWI, which was originally named as venography.[93,94] SWI is reconstructed using single- or multi-echo gradient echo (GRE) magnitude and phase images. The method enhances the visualization of tissues with different magnetic susceptibilities, making it particularly adept at highlighting veins, microbleeds, hemorrhages, and thrombosis where a high concentration of deoxyhemoglobin exists[98] (green arrowheads in Fig. 2a). Because microbleeds and hemorrhages are clinically important, the method has become a routine and invaluable protocol in many neuro exams.[99] Recently, an advanced susceptibility imaging method, phase difference enhanced imaging,[100] has been proposed to enhance different tissues, visualizing fiber tracts such as optic radiation and enhancing the identification of differences in myelin density.[101]

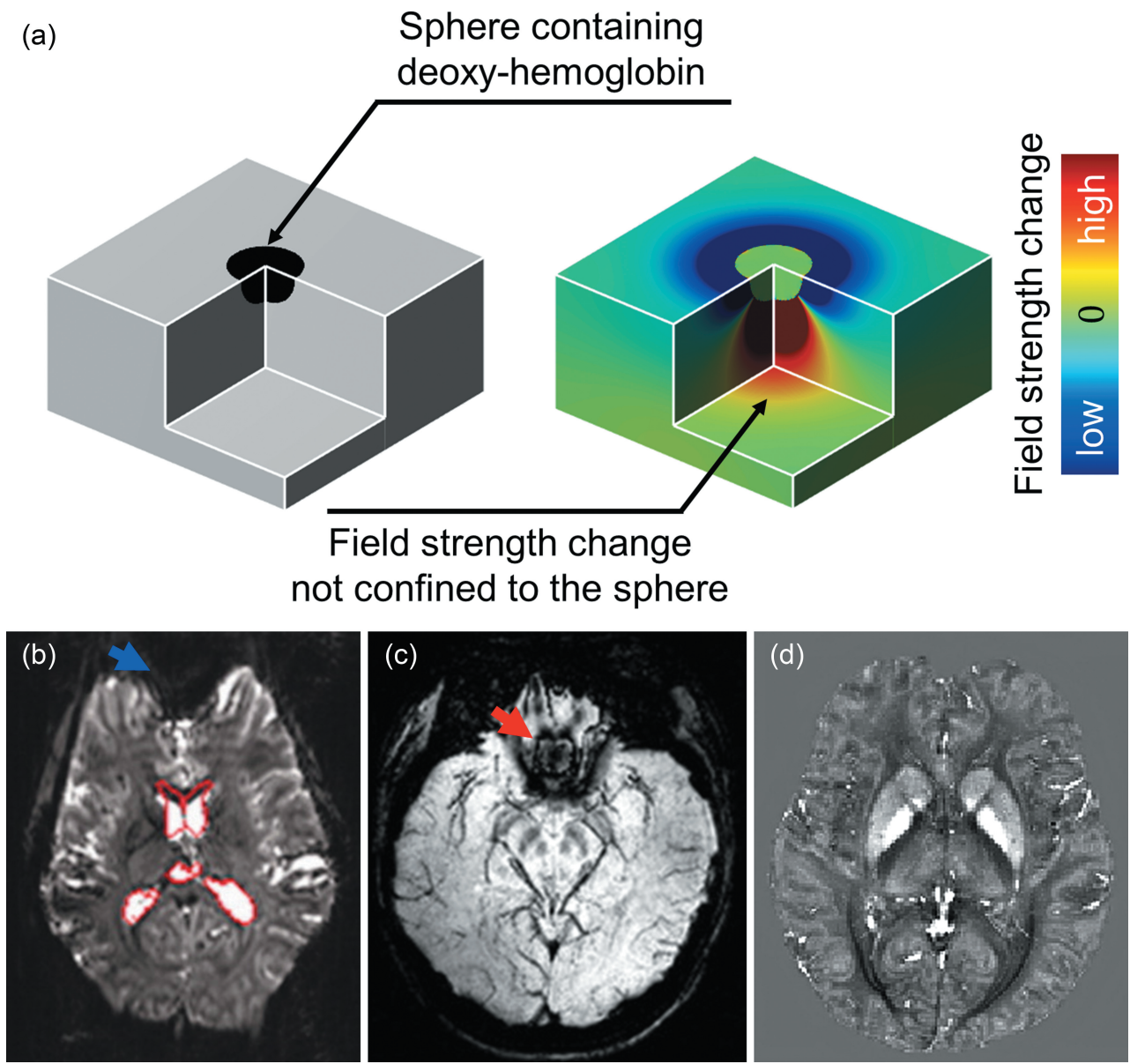

**Fig. 1** (**a**) A demonstration of field perturbation from a paramagnetic sphere in MRI. The sphere is assumed to contain a large number of deoxygenated hemoglobin proteins as the susceptibility source (left), and they create B0 field perturbation in and outside of the sphere (right). (**b**) Susceptibility-induced geometric distortion (blue arrow) in an echo-planar image due to the susceptibility difference between air–tissue interfaces. *Copyright 2011 Wiley. Adapted and used with permission from Se-Hong Oh, Jun-Young Chung, Myung-Ho In, Maxim Zaitsev, Young-Bo Kim, Oliver Speck, and Zang-Hee Cho, Distortion correction in EPI at ultra-high-field MRI using PSF mapping with optimal combination of shift detection dimension, Magnetic Resonance in Medicine, Wiley*. (**c**) An SWI image and its artifact (red arrow). *Copyright 2014 Wiley. Adapted and used with permission from Sung Suk Oh, Se-Hong Oh, Yoonho Nam, Dongyeob Han, Randall B. Stafford, Jinyoung Hwang, Dong-Hyun Kim, HyunWook Park, and Jongho Lee, Improved susceptibility weighted imaging method using multi-echo acquisition, Magnetic Resonance in Medicine, Wiley*. (**d**) A QSM image. *Adapted and reprinted from NeuroImage, 179, Jaeyeon Yoon, Enhao Gong, Itthi Chatnuntawech, Berkin Bilgic, Jingu Lee, Woojin Jung, Jingyu Ko, Hosan Jung, Kawin Setsompop, Greg Zaharchuk, Eung Yeop Kim, John Pauly, and Jongho Lee, Quantitative susceptibility mapping using deep neural network: QSMnet, 119-206, Copyright 2018, with permission from Elsevier*. SWI, susceptibility weighted imaging; QSM, quantitative susceptibility mapping.

## QSM

Unlike traditional MRI, which primarily generates qualitative images, QSM provides quantitative information about the magnetic susceptibility of tissues, offering valuable insights into their microstructural composition.[19] The method utilizes multi-echo GRE phase images to reconstruct quantitative susceptibility maps. Its ability to quantify tissue magnetic susceptibility provides a unique opportunity to differentiate between iron and calcium lesions, a capability derived from the inherent sign difference between diamagnetic calcium and paramagnetic iron, facilitating more accurate diagnoses.[102] In MS, QSM has proven instrumental in categorizing lesions based on their iron and myelin concentration changes, offering insights into disease progression.[103–106] Moreover, QSM plays a pivotal role in elucidating the accumulation of iron in deep gray matter structures, providing valuable information for various neurological disorders, including Parkinson's disease[107,108] (Fig. 2b) and Alzheimer's disease.[109–111] These applications underscore QSM's potential to significantly impact the diagnosis, treatment planning, and understanding of diverse neurological conditions. An advanced QSM method, susceptibility tensor imaging,[112] extends the concept of QSM by considering the anisotropic nature of susceptibility in white matter due to its susceptibility anisotropy,[113] creating fiber orientation information in white matter.

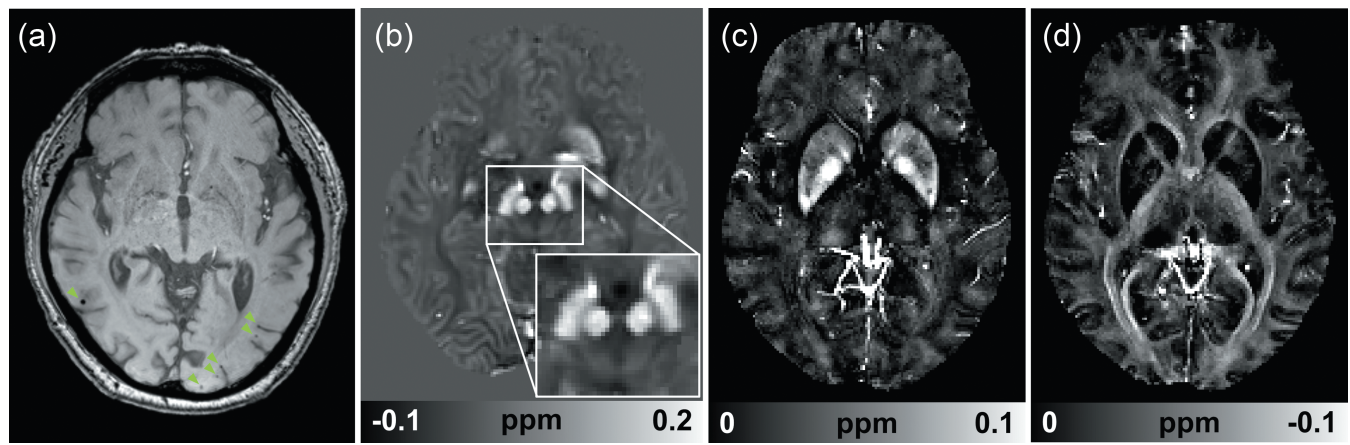

**Fig. 2** (**a**) An SWI image with small hemorrhagic lesions (green arrowheads), (**b**) a QSM map from a Parkinson's disease patient, (**c**) a χ-separation paramagnetic susceptibility map, and (**d**) a diamagnetic susceptibility map. *(**c**) and (**d**) are adapted from NeuroImage, 240, Shin Hyeong-Geol, Lee Jingu, Yun Young Hyun, Yoo Seong Ho, Jang Jinhee, Oh Se-Hong, Nam Yoonho, Jung Sehoon, Kim Sunhye, Masaki Fukunaga, Kim Woojun, Choi Hyung Jin, and Lee Jongho, χ-separation: Magnetic susceptibility source separation toward iron and myelin mapping in the brain, 118,371, Copyright 2021, with permission from Elsevier.* SWI, susceptibility weighted imaging; QSM, quantitative susceptibility mapping.

### Susceptibility source separation

Recently, susceptibility imaging has witnessed significant advancements with the introduction of a susceptibility source separation method (χ-separation or chi-separation or x-separation),[73] a technique that holds great promise in disentangling the mixture of diamagnetic (or negative) sources and paramagnetic (or positive) sources within a voxel. Leveraging a biophysical model of the susceptibility-induced magnetic field perturbation and $R_2'$ (= $R_2^*$ - $R_2$ = $1/T_2^*$ - $1/T_2$), χ-separation allows for the differentiation of dia- and paramagnetic susceptibility sources, assuming a static dephasing regime with the same susceptibility characteristics for both sources (see Fig. 3 for the details of the χ-separation model).[73,75,76,84] This approach enables the creation of separate paramagnetic susceptibility maps and diamagnetic susceptibility maps (Fig. 2c and 2d). In these maps, the definition of dia- and paramagnetic sources is referenced with respect to water. In the brain, where iron constitutes a primary source of paramagnetism and myelin represents a significant diamagnetic source, this technique may deliver iron and myelin distributions of the brain, enabling researchers and clinicians to explore the intricacies of the brain and gain a more nuanced understanding of the pathophysiological processes of diseases (see Question and Answer 9 for current and future applications). However, it is essential to note that χ-separation maps may exhibit inaccuracies stemming from various sources, such as the disruption of the static dephasing regime in regions with high susceptibility concentrations and differences in susceptibility characteristics between ferritin, the primary iron protein in the brain, and myelin, which introduces susceptibility anisotropy (see Questions and Answers 3 and 5).

After χ-separation, more susceptibility source separation methods including DECOMPOSE,[74] χ-sepnet,[114] $R_2^*$QSM,[77,78] and APART-QSM[81,82] have been developed to separate dia- and paramagnetic sources within a voxel. While the original χ-separation utilizes multi-echo GRE images and multi-echo spin echo (SE) images for reconstruction,[73] more recent methods leverage $R_2^*$ instead of $R_2'$, enhancing usability (i.e., $R_2$ map not required).[74,77,114] For example, Chen et al.[74] utilized a different biophysical model that describes a direct relationship between $R_2^*$ and susceptibility sources. In the work of Dimov et al., a linear relationship between $R_2'$ and $R_2^*$ is assumed,[77,78] utilizing a linearly scaled $R_2^*$ map as the input for χ-separation, removing the need for an $R_2$ map. For a similar purpose, a neural network, χ-sepnet-$R_2^*$,[114] is trained to directly reconstruct χ-separation maps from $R_2^*$ and local field inputs.

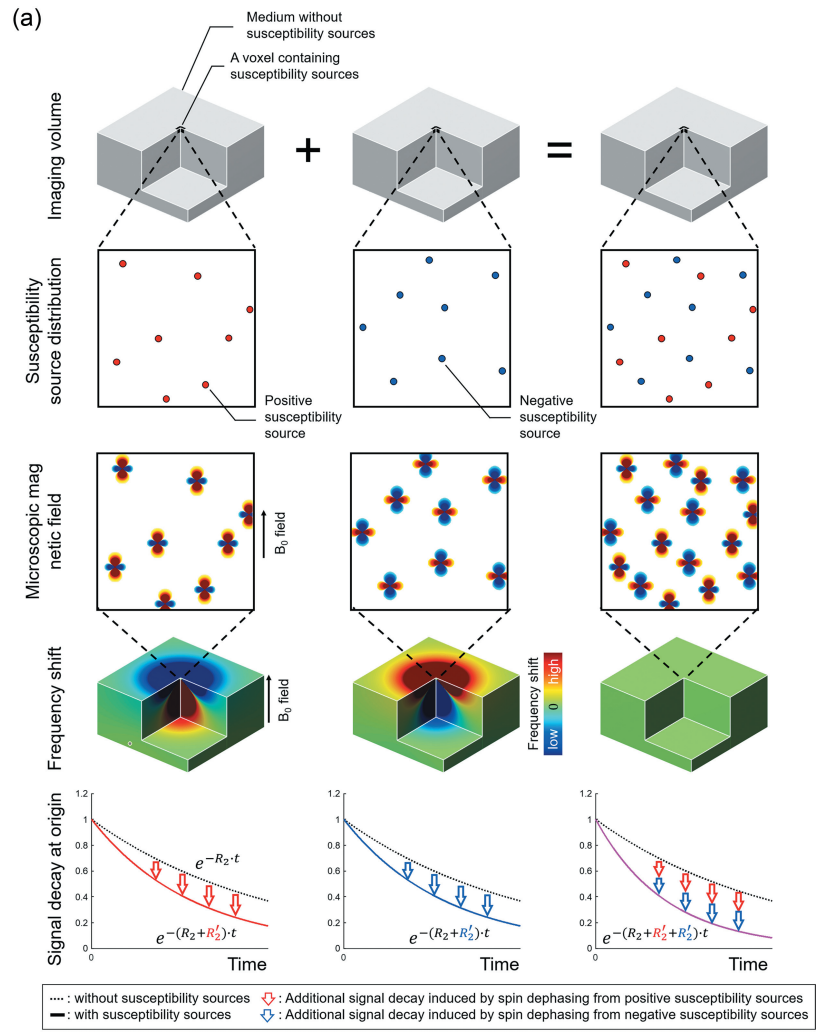

**Fig. 3** Contrast mechanism of χ-separation. (**a**) The effects of para- and diamagnetic susceptibility on frequency and magnitude of MR images. *Adapted and reprinted from NeuroImage, 240, Shin Hyeong-Geol, Lee Jingu, Yun Young Hyun, Yoo Seong Ho, Jang Jinhee, Oh Se-Hong, Nam Yoonho, Jung Sehoon, Kim Sunhye, Masaki Fukunaga, Kim Woojun, Choi Hyung Jin, and Lee Jongho, χ-separation: Magnetic susceptibility source separation toward iron and myelin mapping in the brain, 118,371, Copyright 2021, with permission from Elsevier.* An imaging volume contains randomly distributed susceptibility sources in the voxel at the origin. The first column contains only paramagnetic (or positive) sources, the second column contains diamagnetic (or negative) susceptibility sources, and the third column is composed of para- and diamagnetic sources. The susceptibility sources induce magnetic field perturbation when B0 field is applied, making a frequency shift in the imaging volume. Note that the frequency shift is zero when the same amount of para- and diamagnetic susceptibility sources exist. Inside the voxel containing the susceptibility sources, transverse signal decay with irreversible ($R_2$) and reversible ($R_2'$) decay occurs. Here, when both para- and diamagnetic susceptibility sources exist (3rd column), $R_2'$ is the sum of the $R_2'$ of the para- and diamagnetic sources. (**b**) The frequency domain equation, magnitude domain equation, and combined model for χ-separation. $D_f$ is the field perturbation kernel, and $D_{r,para}(r)$ and $D_{r,dia}(r)$ are the spatially varying relaxometric constants between $R_2'$ and paramagnetic susceptibility, and between $R_2'$ and diamagnetic susceptibility, respectively. The symbol * stands for convolution. Combining the magnitude and phase domain models, $\chi_{para}$ and $\chi_{dia}$ values are calculated by iteratively solving the minimization problem. In the current implementation of χ-separation, a nominal $D_r$ is determined by linear fitting in deep gray matter ROIs. The same $D_r$ is used for both sources disregarding the difference between $D_{r,para}(r)$ and $D_{r,dia}(r)$, and the spatial distribution of them. For the reduction of streaking artifacts, optional regularization terms may be used, similar to QSM. QSM, quantitative susceptibility mapping.

**Question 3:** Can you tell me more about the diamagnetic susceptibility map generated from the susceptibility source separation technique?

**Answer 3:** In the brain, the main diamagnetic susceptibility source is myelin, while other sources, including calcium and some proteins, also exist. Calcium tends to concentrate in focal locations, making it easily distinguishable from myelin. Proteins exhibit varying susceptibilities; for example, oxyhemoglobin is slightly diamagnetic, while deoxyhemoglobin is paramagnetic. However, most proteins, excluding those containing metallic ions like ferritin, generally have susceptibilities close to water. Therefore, myelin stands out as the primary source of the diamagnetic susceptibility map both in gray and white matter. By the way, the major paramagnetic susceptibility source is iron in the form of deoxyhemoglobin, ferritin, and hemosiderin, with some rare diseases introducing additional sources such as copper in Wilson's disease.[115]

Unfortunately, myelin poses unique challenges due to its intricate structure and its complex nature in terms of susceptibility. Myelin exhibits susceptibility anisotropy, meaning its susceptibility measurement varies based on the orientation of the lipid bilayer to B0 or, for myelinated axonal fibers, the fiber orientation to B0.[112,113] Additionally, its multi-compartment microstructure (e.g., axonal space, myelin sheath, and extracellular space) creates a microstructure-induced field perturbation that is also orientation dependent.[116–119] Moreover, not only field perturbation but also $R_2^*$, and consequently $R_2'$, are known to be fiber orientation dependent relative to B0.[120,121]

These characteristics of myelin are not fully accounted for in the current models of susceptibility source separation, assuming the same susceptibility characteristics for both para- and diamagnetic susceptibility sources. As a result, susceptibility maps may contain errors in the quantification of myelin (and iron) concentration. While the effects of the orientation dependence require further investigation, they are anticipated to be relatively small compared to isotropic susceptibility (magnetic susceptibility anisotropy = 0.010 ppm vs. mean magnetic susceptibility = −0.042 ppm reported in reference 122). Thus, the errors from the orientation effects are expected to be limited, and the diamagnetic map from χ-separation effectively reflects myelin distribution in the brain (Fig. 4). We must reemphasize that the diamagnetic susceptibility map is a biomarker of myelin that indirectly reports myelin concentration. It should not be equated to a myelin susceptibility map or a myelin density map.

Aside from the complex issue of myelin, diamagnetic susceptibility maps still contain non-myelin sources (see Question and Answer 5). Therefore, researchers should be careful in interpreting the maps.

**Question 4:** What is the evidence that the diamagnetic susceptibility map reflects myelin distribution?

**Answer 4:** Several findings suggest a strong association between the diamagnetic susceptibility map and myelin distribution. However, before delving into these findings, we want to emphasize that, just like other myelin imaging methods, the diamagnetic susceptibility map does not "equate" to a myelin map (see Questions and Answers 3 and 5). Nevertheless, we believe it serves as a valuable surrogate biomarker for myelin, holding significant potential for various applications.

The first evidence that demonstrates the relationship between the diamagnetic susceptibility map and myelin distribution is the results from an *ex-vivo* brain specimen that includes the primary visual cortex (Fig. 5[73]). When the susceptibility map and myelin histology via Luxol fast blue (LFB) are compared, they show qualitatively similar distributions. For example, both images show a well-known cortical laminar structure in the primary visual cortex (yellow arrows: line of Gennari) and consistent contrasts among the three white matter fibers (red triangle: optic radiation; purple square: stratum sagittale internum; and green stars: forceps). When we reanalyzed the data for quantitative comparison, a high correlation ($R^2 = 0.770$; Fig. 5d) was measured between the negative susceptibility values and LFB optical density (OD) values in the ROIs (yellow circles and rectangles in Fig. 5c), demonstrating that the diamagnetic susceptibility reports myelin concentration. When the cortex of the primary visual cortex (cyan region in Fig. 5c) was segmented to generate a cortical profile, the results also confirm a strong consistency between the two profiles (Fig. 5e), validating the technique for cortical myelin assessment. Both profiles successfully reveal the line of Gennari (green arrow in Fig. 5e). These results suggest that we can generate a cortical profile of myelin using susceptibility source separation, opening the potential of exploring myelin concentration in the cortex, as recently demonstrated in *in-vivo* human brain at 3T.[35]

Another evidence from an *ex-vivo* specimen of the macaque brain can be found in the work by Li et al.,[81] demonstrating a strong correlation between the diamagnetic susceptibility values and normalized myelin contents from LFB-stained images among white matter fibers ($R^2 = 0.530$ in χ-separation and $R^2 = 0.854$ in APART-QSM).

In MS lesions, a study of $R_2^*$QSM-based susceptibility source separation method demonstrated that the diamagnetic susceptibility maps of MS tissue samples report a reasonable correlation (correlation coefficient = 0.47, ROI analysis) with the OD of myelin basic protein antibody stain.[78] This finding indicates that the map does show the connection to myelin even in pathological conditions, suggesting possibilities toward exploring brain pathologies using the technique.

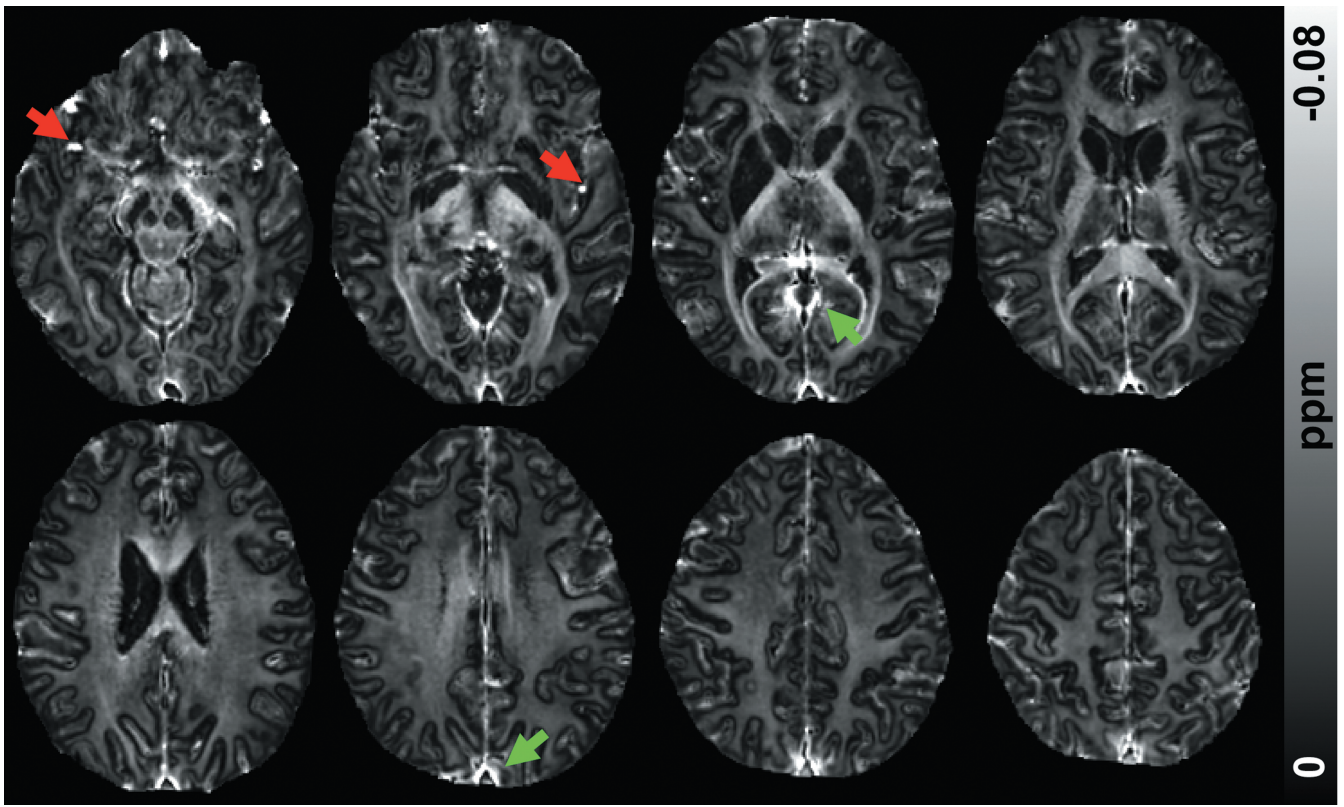

**Fig. 4** An example of a diamagnetic susceptibility map from a representative subject. Data were acquired at 3T with 0.75 mm isotropic resolution and FOV 192 × 167 × 120 mm$^3$ during the acquisition time of 5 min 51 sec. The diamagnetic susceptibility map was reconstructed using χ-sepnet-$R_2$*, which utilizes $R_2$* instead of $R_2'$. The red arrows display the vessel-related artifacts, and the green arrows display the non-local $R_2$*-related artifacts.

Note that all of these validations are from *ex-vivo* samples, which do not contain vessels or other sources of artifacts observed *in vivo*. Hence, further complications may exist in *in-vivo* results.

For *in-vivo* maps, one can check the atlas of the diamagnetic susceptibility maps averaged over a large number of individuals.[82,83] Overall, well-known fibers that connect long distances in the brain (e.g., corpus callosum, optic radiation, and internal capsule) reveal higher absolute contrast values ($|\chi_{dia}|$, Fig. 3 in reference 83). Of course, this contrast distribution can be observed in individual subjects as well (Fig. 4). When the atlas is compared with that of MWI, they show pretty good correspondence, reporting a correlation of $R^2 = 0.63$ in the ROIs of white matter (Fig. 7 in reference 83). This result further corroborates that the diamagnetic susceptibility maps reflect myelin distribution in white matter although methodology-specific differences appear in the atlases (e.g., vessels and fiber orientation-dependent contrast variations in the diamagnetic susceptibility atlas; overestimation of myelin water fraction in internal capsule in the MWI atlas).

Hence, both *ex-vivo* and *in-vivo* results confirm that the diamagnetic susceptibility is well-correlated with myelin distribution, suggesting that it can be used as an imaging biomarker for myelin.

**Question 5: I can see non-myelin structures in the diamagnetic susceptibility map. What are they?**

**Answer 5:** A diamagnetic susceptibility map reconstructed from a susceptibility source separation method may contain non-myelin structures that complicate the direct interpretation of the map as a myelin distribution map. In the χ-separation maps, the most conspicuous source of artifacts is large vessels (Fig. 4, marked with red arrows). Flow inside the vessels induces signal variation and spatial displacement, resulting in inconsistent signal decay across TE in the voxels inside and near the vessels. This inconsistency leads to incorrect $R_2$* values, thus introducing errors in the resulting diamagnetic susceptibility values. Furthermore, the voxels inside the vessels violate the assumption of the static dephasing regime, further complicating the problem.

Another source of artifacts is non-local $R_2$* effects due to large susceptibility differences at air–tissue interfaces and around large veins (Fig. 4, marked with green arrows). It can induce overestimation of $R_2^*$ values, producing errors in the diamagnetic susceptibility map. One may correct for the non-local $R_2^*$ effects based on signal models.[123,124]

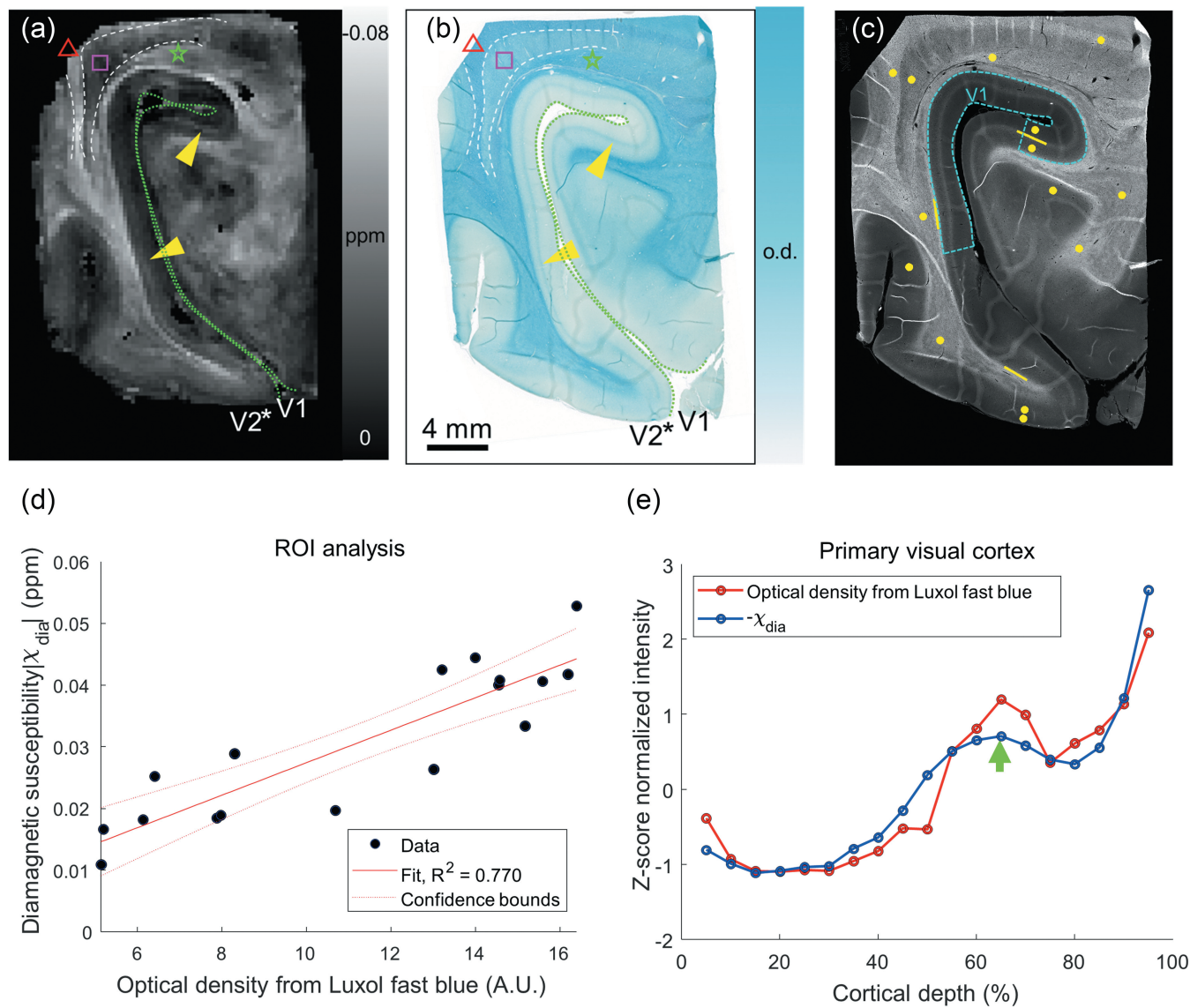

**Fig. 5** (**a**) A diamagnetic susceptibility map from χ-separation and (**b**) an LFB-stained image. (*a*) and (*b*) are adapted and reprinted from *NeuroImage*, 240, Shin Hyeong-Geol, Lee Jingu, Yun Young Hyun, Yoo Seong Ho, Jang Jinhee, Oh Se-Hong, Nam Yoonho, Jung Sehoon, Kim Sunhye, Masaki Fukunaga, Kim Woojun, Choi Hyung Jin, and Lee Jongho, χ-separation: Magnetic susceptibility source separation toward iron and myelin mapping in the brain, 118,371, Copyright 2021, with permission from Elsevier. (**c**) The LFB-stained image is converted to an LFB OD image for quantitative analysis. For an ROI analysis, 17 ROIs were chosen and drawn on each of the LFB OD and the diamagnetic susceptibility map. The ROIs are displayed as yellow circles and lines. (**d**) The ROI analysis result reveals good correlation between LFB OD and the diamagnetic susceptibility ($R^2 = 0.770$). (**e**) The cortical profiles of the diamagnetic susceptibility ($-\chi_{dia}$) and LFB OD from the primary visual cortex ROI (cyan dashed line in (**c**)). The two profiles agree, revealing the well-known mid-cortical laminar structure of the line of Gennari (green arrow). LFB, luxol fast blue; OD, optical density.

Similar to QSM, streaking artifacts from imperfect dipole deconvolution are also a source of error. When using regularization-based algorithms (e.g., morphology enabled dipole inversion; MEDI[125]) for reconstruction, streaking artifacts may introduce erroneous estimation of myelin concentration. This can be mitigated by using multi-orientation data, as in the calculation of susceptibility through multiple orientations sampling (COSMOS)[126] reconstruction in QSM, but is not practical because multiple scans are necessary, elongating the scan time. Reconstruction using a neural network trained with multi-orientation reconstructed data (e.g., χ-sepnet[114]) can provide streaking artifact-free images (Fig. 4) as demonstrated in QSM.[97]

**Question 6: How do I acquire data for susceptibility source separation?**

**Answer 6:** The original χ-separation method involves acquiring 3D multi-echo GRE for the field and $R_2^*$ maps and 2D multi-echo SE for the $R_2$ map. After the acquisition, the field and $R_2'$ maps, where $R_2'$ is calculated by $R_2^*-R_2$, are fed into the χ-separation algorithm.

For the multi-echo GRE protocol, you can use the recent recommendation outlined in the QSM consensus paper, which suggests acquiring the whole brain in 6 min of scan time at the resolution of 1 mm isotropic voxel size with 5 echoes at 3T. For the SE protocol, multiple options exist; for example, a custom-designed 2D multi-echo SE with 6 echoes, an in-plane resolution of $1 \times 1$ mm$^2$ and a slice thickness of 2 mm at the scan time of 12.4 min was used in the original χ-separation paper.[73] To shorten the scan time, a 2D dual echo Turbo SE product sequence can be employed, with parameters such as TR = 11000 ms, TE = 10 and 100 ms, in-plane resolution = $0.6 \times 0.6$ mm$^2$ (required for clinical evaluation and later interpolated to 1 mm$^2$ for χ-separation), slice thickness = 2 mm, turbo factor = 7, acceleration factor = 2, and scan time = 7 min. In both cases, the thicker slice in the SE data needs to be interpolated to 1 mm to match the resolution of the GRE data. Advanced acquisition schemes have been explored for rapid simultaneous acquisition of GRE and SE data.[79,127]

However, it is worth to note that all the SE acquisition methods or simultaneous acquisition approaches mentioned above may not be used in routine clinical scans or may not be available as a product sequence. This can result in substantially increased scan time, limiting the applicability of χ-separation. To overcome this challenge, alternative methods such as linear scaling of $R_2^*$ to $R_2'$,[77,78,82] deep learning-powered $R_2'$ generation,[64,114] and a new model that only relies on multi-echo GRE data[74] have been proposed. While these methods improve the applicability, they may come with a trade-off of compromised accuracy. A few comparison results are underway,[77,128] and further research is required to fully evaluate and compare different methods, particularly in clinical use cases.

**Question 7: What about data processing? Any processing tools available for susceptibility source separation?**

**Answer 7:** Yes, χ-separation toolbox is available (https://github.com/SNU-LIST/chi-separation), which includes the processing for the original χ-separation with the frequency and $R_2'$ input, as well as deep learning-powered χ-separation using the frequency and $R_2^*$ input or frequency and $R_2'$ input. The toolbox incorporates all the pre-processing steps required for χ-separation, offering a comprehensive solution with multiple options, including denoising, for data processing. Other toolboxes are also accessible (https://github.com/AMRI-Lab/APART-QSM), providing alternatives for data analysis in susceptibility source separation techniques.

Here are additional details about data processing in susceptibility source separation, which requires generating local field and $R_2'$ (or $R_2^*$) maps as the input to the source separation algorithm.

### Local field
Local field perturbation from microstructural susceptibility sources is estimated similarly to QSM. Briefly, phase images from the multi-echo GRE data are unwrapped and echo combined (or echo combined and then unwrapped). Subsequently, a background field is removed to generate a local field perturbation map. We highly recommend readers to review the recent QSM consensus paper to understand each step.

### $R_2$ and $R_2^*$
An $R_2'$ map is generated by the subtraction of $R_2$ from $R_2^*$. This process requires generation of $R_2$ and $R_2^*$ maps from multi-echo SE and multi-echo GRE, respectively. While creating an $R_2^*$ map can be straightforward by fitting an exponential decay function to the multi-echo magnitude data in each voxel, this map often becomes noisy and serves as a primary source of noise in χ-separation maps. Hence, denoising the map via principal component analysis or deep learning may be useful.[129] Additionally, a few voxels may contain errors due to the background field or rapid loss of signal, requiring careful processing. Generating an $R_2$ map is more complicated due to the stimulated echoes in multi-echo SE.[130] This effect is well-known but can be a surprise for many people. It can be identified with the signature characteristics of the second echo reporting higher signal intensity than the first echo, necessitating correction methods for accurate $R_2$ estimation.[130] Note that this issue does not happen in single echo SE and is largely mitigated in Turbo SE.

### Susceptibility source separation algorithm
Once datasets are preprocessed, one can run a susceptibility source separation algorithm to generate para- and diamagnetic susceptibility maps. As mentioned in the previous section, several methods exist, and we will explain χ-separation as an example. The algorithm estimates the para- and diamagnetic susceptibility concentrations by iteratively solving the minimization problem in Fig. 3b. Similar to QSM, it requires solving the ill-conditioned dipole deconvolution problem, necessitating additional regularization such as MEDI[125] or iterative linear equation and least-squares (iLSQR),[131] which are implemented in the toolbox. One may acquire multi-orientation dataset to convert the ill-conditioned problem to a well-conditioned one, generating a gold-standard result. Alternatively, one can employ a neural network implementation of χ-sepnet-$R_2'$ or χ-sepnet-$R_2^*$, which are trained on maps reconstructed using multi-orientation data.[132] Reconstruction using these networks significantly improves image quality. The χ-separation algorithm has been shown to select a long $T_2$ location (mostly cerebrospinal fluid; CSF) as the reference with zero susceptibility,[133] differentiating it from QSM, which is required to set an explicit reference region (e.g., CSF or whole brain).

**Question 8:** I know there are other myelin imaging methods. What are the pros and cons of the susceptibility source separation compared to them?

**Answer 8:** Several MRI techniques are available for visualizing myelin, either directly or indirectly. Comparing them is a complex task, and we recommend referring to the following review papers.[23–28] In this section, we will briefly introduce well-known methods and then compare them against the susceptibility source separation technique.

MWI involves imaging water protons, known as myelin water, between the lipid bilayers of myelin. This myelin water exhibits a shorter relaxation time ($T_2$, $T_2^*$, or $T_1$) than that of axonal or extracellular water, and this difference is exploited to generate MWI.[1–4] Another approach involves directly visualizing protons bound by the myelin lipid bilayers. These protons have limited mobility and are strongly coupled, leading to a much shorter transverse relaxation time (<<1 ms) to be imaged using a conventional sequence. Studies have developed ultra-short TE (UTE) imaging to address this challenge.[15,17] Alternatively, magnetization transfer effects[5] can be utilized to indirectly generate images from macromolecules. Several methods exist, including magnetization transfer ratio, magnetization transfer saturation, quantitative magnetization transfer,[7] macromolecular proton fraction,[14] and inhomogeneous magnetization transfer.[8] Myelin volume fraction from synthetic MRI[10,11,134] also provides myelin information by relating the simultaneously measured $T_1$, $T_2$, and proton density (PD) values to four partial volume components consisting of myelin, free water, cellular, and excess parenchymal partial volume components. Finally, quantitative $R_1$ imaging,[12,16] $R_2^*$ imaging,[135] and $T_1$-weighted over $T_2$-weighted imaging[13] have also been proposed as indirect approaches to estimate myelin concentration in the brain.

All of these methods have their cons and pros. When comparing these methods, the susceptibility source separation technique with $R_2^*$-only (using only GRE data) has two clear advantages: acquisition in high resolution and application in ultra-high-field MRI. As demonstrated in Fig. 4, we can generate a 0.75-mm isotropic resolution map even at 3T in less than 6 min of scan time. At an ultra-high-field strength, the technique benefits from both increased signal and increased susceptibility effects, creating a high-quality map at a resolution of 0.6 mm at 7T.[136] This advantage can be further extended to the susceptibility source separation method using $R_2'$ when advanced acquisition schemes that allow simultaneous and rapid acquisition of GRE and SE data are applied.[79] On the other hand, MWI and UTE imaging suffer from a low SNR, limiting their resolution. The synthetic MRI approach is limited by spatial resolution. Magnetization transfer requires a high specific absorption rate, which can restrict its application in ultra-high fields. Other methods ($R_1$, $R_2^*$, and $T_1$-weighted over $T_2$-weighted) are unable to remove contributions of iron, providing limited specificity (e.g., see Fig. 4 in reference 35). When it comes to the disadvantage of the susceptibility source separation, it shares the well-known challenge of GRE acquisition including sensitivity to spatial and temporal B0 field inhomogeneity. In contrast, myelin imaging methods such as SE-based MWI,[1] UTE,[15,17] and myelin volume fraction from synthetic MRI[11] are relatively robust to B0 field inhomogeneity. Additionally, all the non-myelin structures addressed in Question and Answer 5 may complicate the direct interpretation of the map as myelin.

**Question 9:** Applications so far and for the future?

**Answer 9:** Susceptibility source separation methods have found valuable applications in various neuroimaging studies, holding significant implications for both current clinical practices and future research endeavors. In MS studies, these techniques have proven instrumental in identifying lesions (Fig. 6), particularly in cases with decreased iron and decreased myelin concentrations that may go unnoticed in conventional QSM (lesion 2 in Fig. 6). These changes often lead to minimal alteration in QSM values, rendering these lesions potentially unidentified by QSM alone. However, susceptibility source separation provides sensitivity for detecting such lesions. Additionally, these methods showcase potential in comparing lesion characteristics between MS and Neuromyelitis Optica, offering a new avenue for differentiating these two diseases based on their lesion features.[64] The applications of susceptibility source separation are poised to extend to diagnosis, disease progression monitoring, and treatment efficacy assessment, with a notable potential to report on remyelination processes within lesions.

Beyond well-known demyelination diseases, susceptibility source separation methods may demonstrate broader applications. In neurodegenerative disorders such as Alzheimer's disease and Parkinson's disease, these techniques can facilitate the visualization of myelin and iron changes, providing insights into the structural alterations associated with these conditions.[80] The methodology may extend its reach to psychiatric disorders, including schizophrenia, bipolar disorder, and depression, where ongoing research explores the role of myelin.[137] Additionally, in the identification of seizure locations or in evaluating traumatic brain injury, the method may prove useful.[138,139] Furthermore, susceptibility source separation can be applied in neuroscience, exploring developmental and aging brains. The study of myeloarchitecture can provide valuable insights into brain parcellation.[35] The current and potential applications of susceptibility source separation underscore its transformative potential in advancing our understanding of neurological and psychiatric conditions, as well as its contributions to neuroscience.

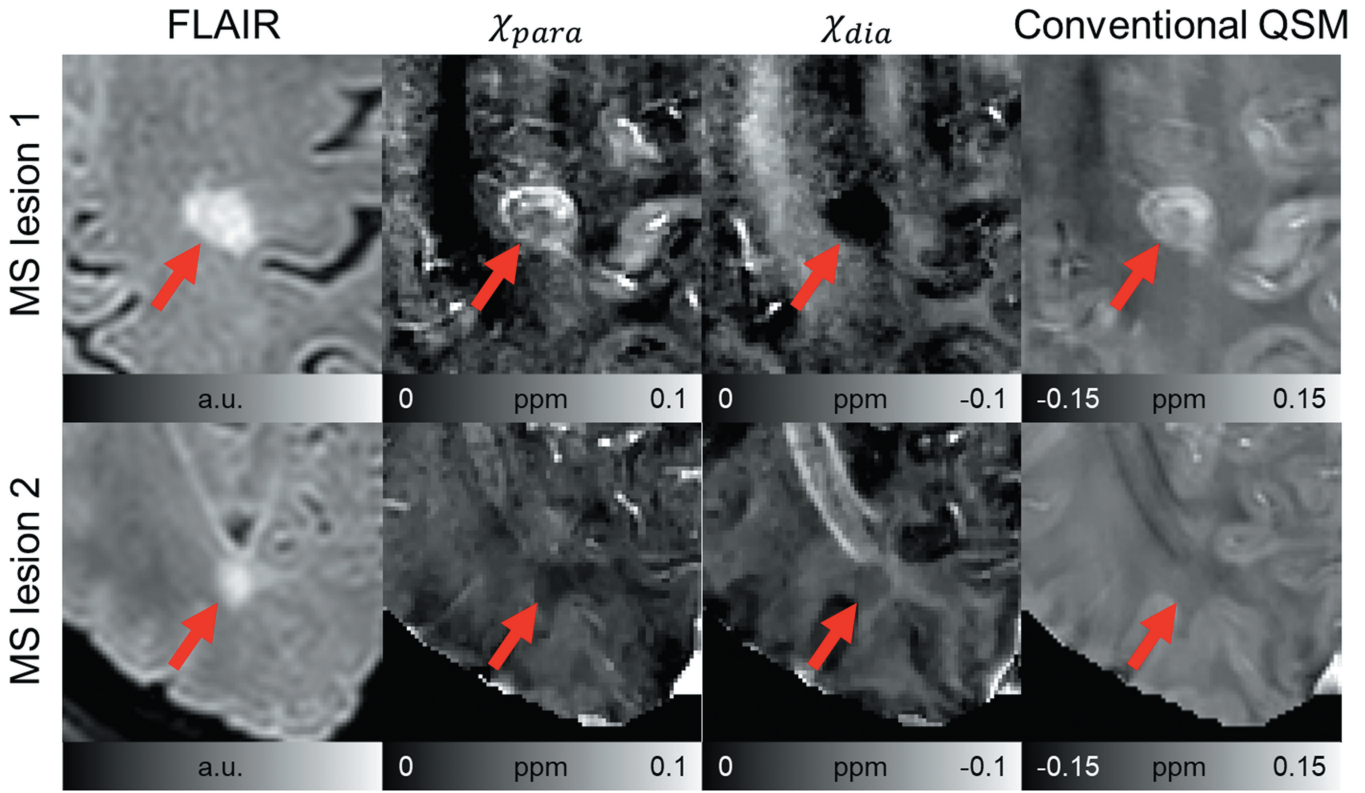

**Fig. 6** The *in-vivo* results of two characteristic MS lesions. *Adapted and reprinted from NeuroImage, 240, Shin Hyeong-Geol, Lee Jingu, Yun Young Hyun, Yoo Seong Ho, Jang Jinhee, Oh Se-Hong, Nam Yoonho, Jung Sehoon, Kim Sunhye, Masaki Fukunaga, Kim Woojun, Choi Hyung Jin, and Lee Jongho, χ-separation: Magnetic susceptibility source separation toward iron and myelin mapping in the brain, 118,371, Copyright 2021, with permission from Elsevier.* The red arrows indicate the MS lesions identified in the FLAIR images. The MS lesion in the first row displays a paramagnetic rim sign on both $\chi_{para}$ and conventional QSM maps. The $\chi_{dia}$ map suggests demyelination along with iron accumulation in the lesion, which cannot be identified using the conventional QSM. On the second row, the MS lesion displays low $\chi_{para}$ and $\chi_{dia}$ values, suggesting decreased iron and myelin. However, this lesion cannot be identified in the QSM map because both paramagnetic and diamagnetic susceptibility values have decreased, leading to isointense lesion in the QSM map. MS, multiple sclerosis; FLAIR, fluid-attenuated inversion recovery; QSM, quantitative susceptibility mapping.

## Conclusion

In conclusion, this review has traversed the exciting landscape of the rapidly growing area of the magnetic susceptibility source separation technique as a promising biomarker for myelin. As we stand at the forefront of this burgeoning field, we anticipate continued technical development, further validation, and the exploration of new applications, particularly capitalizing on its powerful advantage in high-resolution imaging at high-field MRI. It is also important to recognize that this technique extends beyond myelin, providing iron distribution, thus offering a complementary dimension to our understanding of the brain. However, it remains imperative to acknowledge that, like any method, magnetic susceptibility source separation has its limitations. A negative susceptibility map, while informative, does not equate to myelin content, emphasizing the importance of careful interpretation, particularly in pathological conditions. We strongly recommend that investigators intending to apply this technique understand the details of the technology, and we hope this review serves as a guiding step in that direction. Simultaneously, developers are encouraged to invest efforts in refining methods that allow for the direct interpretation of biological information, further propelling the potential of magnetic susceptibility source separation in advancing our understanding of the brain.

## Acknowledgments

This work was supported by the National Research Foundation of Korea (NRF-2021R1A2B5B03002783 and NRF-2019M3C7A1031994), IITP-2023-RS-2023-00256081,